\documentclass[conference, 10pt]{IEEEtran}

\usepackage[style=ieee, backend=biber, isbn=false]{biblatex}

\newif\ifshowinfo
\showinfotrue

\usepackage{amsmath}

\usepackage{url}
\usepackage{graphicx}
\usepackage{color}
\usepackage{placeins}
\usepackage{float}
\usepackage{hyperref}
\usepackage{tabularx,colortbl}
\usepackage{ifthen}

\usepackage{mathtools}
\usepackage{siunitx}
\usepackage{comment}
\usepackage{bm}
\usepackage{cleveref}
\newcommand{\mathbfit}[1]{\bm{#1}}
\newcommand{\mat}[1]{\mathbf{#1}}

\makeatletter

\newcounter{authcount}
\renewcommand{\author}[2][]{
   \stepcounter{authcount}
   \@namedef{author@\theauthcount}{#2}
   \@namedef{authorlabel@\theauthcount}{#1}
}

\newcounter{addrcount}
\newcommand{\address}[2][]{
   \stepcounter{addrcount}
   \@namedef{address@\theaddrcount}{#2}
   \@namedef{addresslabel@\theaddrcount}{#1}
}

\newcommand{\alsep}{and}

\def\newmaketitle{\par%
  \begingroup%
  \normalfont%
  \def\thefootnote{}
  \def\footnotemark{}
  \let\@makefnmark\relax
  \footnotesize
  \footnotesep 0.7\baselineskip
  \normalsize%
  \twocolumn[\thenewmaketitle\@IEEEaftertitletext]%
  
  \if@IEEEusingpubid
     \enlargethispage{-\@IEEEpubidpullup}%
  \fi
  \endgroup
  \setcounter{footnote}{0}\let\maketitle\relax\let\@maketitle\relax
  \gdef\@thanks{}%
  
  \let\thanks\relax}

\def\thenewmaketitle{

  \newpage
  \begin{center}%
    \vskip0.2em{\Huge\@IEEEcompsoconly{\sffamily}\@IEEEcompsocconfonly{\normalfont\normalsize\vskip 2\@IEEEnormalsizeunitybaselineskip
   \bfseries\large}\@title\par}\vskip1.0em\par%
    \vspace{1ex}
    \newcounter{authloop}
    \newcounter{c@tmp}
    \ifthenelse{\value{authcount}=2}{%
      \newcommand{\liand}{ and }}{%
      \newcommand{\liand}{, and }}
  
    \ifthenelse{\value{addrcount}<2}{%
      
      \@nameuse{author@1}%
      \stepcounter{authloop}%
      \whiledo{\value{authloop}<\value{authcount}}{%
        \setcounter{c@tmp}{\value{authcount}}%
        \addtocounter{c@tmp}{-\value{authloop}}%
        \ifthenelse{\value{c@tmp}=1}{%
          \renewcommand{\alsep}{\liand}}{\renewcommand{\alsep}{, }}%
        \stepcounter{authloop}\alsep \@nameuse{author@\theauthloop}}\\%
    }
    {
      \@nameuse{author@1}${}^{(\@nameuse{authorlabel@1})}$%
      \stepcounter{authloop}%
      \whiledo{\value{authloop}<\value{authcount}}{%
      \setcounter{c@tmp}{\value{authcount}}%
      \addtocounter{c@tmp}{-\value{authloop}}%
      \ifthenelse{\value{c@tmp}=1}{%
        \renewcommand{\alsep}{\liand}}{\renewcommand{\alsep}{, }}%
      \stepcounter{authloop}\alsep \@nameuse{author@\theauthloop}%
        ${}^{(\@nameuse{authorlabel@\theauthloop})}$%
      }
    }

    \vspace{0.2ex}

    \ifthenelse{\value{addrcount}>0}{%
      \ifthenelse{\value{addrcount}=1}{
        
        {\@nameuse{address@1}}
      }
      {
        \newcounter{c@address}

        \begin{center}
        \whiledo{\value{c@address}<\value{addrcount}}
        {
          \refstepcounter{c@address}
            ${}^{(\thec@address)}$\,%
              \label{\@nameuse{addresslabel@\thec@address}}%
              \@nameuse{address@\thec@address}\\ %
        }
        \end{center}
      } 
    }
    {
      \relax
    }
  \end{center}
}

\makeatother

\title{High-Frequency Preconditioners for
Electromagnetic Integral Equations Based
on Helmholtz Regularizations}

\ifshowinfo

\author[\ref{org1}, \ref{org2}]{Simone Ciciriello}
\author[\ref{org1}]{Viviana Giunzioni}
\author[\ref{org3}]{Alexandre Dély}
\author[\ref{org4}]{Adrien Merlini}
\author[\ref{org5}]{Simon B. Adrian}
\author[\ref{org1}]{Francesco P. Andriulli}

\address[org1]{Department of Electronics and Telecommunications, Politecnico di Torino, Turin, Italy}
\address[org2]{Early Research Honors School, Politecnico di Torino, Turin, Italy}
\address[org3]{Thales DMS, Elancourt, France}
\address[org4]{Microwave Department, IMT Atlantique, Brest, France}
\address[org5]{Fakultät für Informatik und Elektrotechnik, Universität Rostock, Rostock, Germany}

\fi

\ifshowinfo
   \hypersetup{pdftitle={High-Frequency Preconditioners for Electromagnetic Integral Equations Based on Helmholtz Regularizations}}
\else
   \hypersetup{
       pdfauthor={},
       pdfcreator={},
       pdfproducer={},
       pdftitle={High-Frequency Preconditioners for Electromagnetic Integral Equations Based on Helmholtz Regularizations},
       pdfsubject={}
   }
\fi

\begin{document}

\newmaketitle

\begin{abstract}
The numerical solution of the Electric Field Integral Equation (EFIE) via the Boundary Element Method (BEM) can be computationally challenging due to conditioning issues arising in different regimes, such as (i) when the frequency decreases and the discretization density remains constant, (ii) when the frequency is kept constant while the discretization is refined, and (iii) when the frequency increases along with the discretization density. To address these issues, several preconditioning approaches for the related matrix system have been developed in the literature, only a few of which address all regimes simultaneously.
This paper investigates one of these techniques and presents a strategy for accelerating the associated matrix-vector products (MVPs). 
In particular, we propose a novel preconditioning strategy for the shifted Helmholtz operator, for which standard pseudo-inversion techniques have shown unsatisfactory results. Instead, the application of our preconditioning technique stabilizes the number of iterations in all the aforementioned regimes. In view of these achievements, the pseudo-inversion of the shifted Helmholtz operator can be obtained in quasi-linear complexity when proper acceleration strategies are used, thus enabling the numerical solution of the EFIE with the same complexity.

\end{abstract}

\section{Introduction}
Among the variety of numerical techniques for modeling the scattering of an incident electromagnetic wave by a perfectly electrically conducting (PEC) object, one of the most effective is the Boundary Element Method (BEM) applied to integral equations. The Electric Field Integral Equation (EFIE) is one of the most popular integral equations in this framework \cite{jin2015theory}. Nevertheless, the BEM leads to a dense linear system that is challenging to solve directly due to its large size. Hence, it is often advantageous to resort to iterative methods, such as the Generalized Minimal Residual (GMRES) method, which, however, fail to converge in a limited number of iterations if the system matrix is ill-conditioned, as is the case for the EFIE \cite{adrian2021electromagnetic}. 

The ill-conditioned nature of the system that discretizes the EFIE manifests itself under specific conditions such as (i) when the frequency decreases while keeping the discretization density constant, (ii) when the frequency is kept constant while the discretization is refined, and (iii) when the frequency increases along with the discretization density \cite{adrian2021electromagnetic}. These breakdown phenomena lead to a sharp increase in the number of iterations required by iterative solvers to converge, making large-scale problems practically intractable.

Diverse stabilization strategies have been proposed to tackle one or more of the three breakdowns presented above, but very few of them address all three breakdowns simultaneously. This is the case for the formulation presented in \cite{dely2019preconditioning}, which constitutes the starting point of this work. 
In \cite{dely2019preconditioning} a new preconditioning strategy for the EFIE is proposed, based on the quasi-Helmholtz decomposition of the electric field operator and on the separate regularization of the solenoidal and non-solenoidal components using shifted Helmholtz operators.
While \cite{dely2019preconditioning} proposes an effective stabilization strategy for the EFIE, further technical clarification is required to enable coupling with fast solvers (e.g., the Fast Multipole Method, FMM \cite{jin2015theory}) and the solution of the system with quasi-linear complexity. 

In this work, our aim is to improve the efficiency of the matrix-vector product (MVP) required for the implementation of this formulation. In particular, our focus is on the preconditioning block for the quasi-irrotational component of the EFIE, which requires the pseudo-inversion of a discretized shifted Helmholtz operator.
Unfortunately, the (shifted) Helmholtz operator suffers, similarly to the EFIE, from (ii) the dense-discretization and (iii) the high-frequency breakdown; the application of standard techniques for its efficient pseudo-inversion has shown unsatisfactory results in terms of computational complexity.
To overcome these issues, we propose a new preconditioning strategy for the shifted Helmholtz operator matrix which stabilizes both the matrix condition number and the number of iterations in the three aforementioned regimes for a large class of scatterers.

\section{Background and Notation}
\label{sec:background}

Given a simply connected PEC body with smooth surface $\Gamma$ in a background medium of permittivity $\varepsilon$ and permeability $\mu$,
let $\mathbfit{j}$ denote the surface current density induced on $\Gamma$ by the time-harmonic incident electric field $\mathbf{E}^i$ of frequency $f$. The current density $\mathbfit{j}$ satisfies the EFIE,
\begin{equation}
    \eta (\mathcal{T}_k \mathbfit{j}) (\mathbfit{r}) = -\hat{\mathbf{n}}(\mathbf{r}) \times \mathbf{E}^i (\mathbf{r})
\end{equation}
where $\eta = \sqrt{\mu/\varepsilon}$, $k= 2\pi f \sqrt{\mu \varepsilon}$ and $\hat{\mathbf{n}}$ is the outward normal to $\Gamma$.
The electric field integral operator is defined as
\begin{equation}
    (\mathcal{T}_k \mathbfit{j}) (\mathbfit{r}) = -\mathrm{j}k \, (\mathcal{T}_{s,k} \mathbfit{j}) (\mathbfit{r}) 
    - \frac{1}{-\mathrm{j}k} \, (\mathcal{T}_{h,k} \mathbfit{j}) (\mathbfit{r}) \ ,
\end{equation}
where, given the free-space Green's function 
\begin{equation}
    G_k(\mathbfit{r}, \mathbfit{r'}) = \frac{e^{-\mathrm{j}k|\mathbfit{r}-\mathbfit{r}'|}}{4\pi|\mathbfit{r}-\mathbfit{r}'|},
\end{equation}
the operators $\mathcal{T}_{s,k}$ and $\mathcal{T}_{h,k}$ are defined by
\begin{align}
    (\mathcal{T}_{s,k} \mathbfit{j}) (\mathbfit{r}) &= \hat{\mathbfit{n}}(\mathbfit{r}) \times 
    \int_{\Gamma} G_k(\mathbfit{r}, \mathbfit{r'}) \, \mathbfit{j}(\mathbfit{r}') \, \mathrm{d}S' \ , \\
    (\mathcal{T}_{h,k} \mathbfit{j}) (\mathbfit{r}) &= \hat{\mathbfit{n}}(\mathbfit{r}) \times \nabla 
    \int_{\Gamma} G_k(\mathbfit{r}, \mathbfit{r'}) \, 
    \nabla' \cdot \mathbfit{j}(\mathbfit{r}') \, \mathrm{d}S' \ .
\end{align}

The equation can be numerically solved by applying the BEM, i.e., by approximating the current as a linear combination of $N$ basis functions $\{ \mathbfit{f}_i \}_{i=1}^N$ ($\mathbfit{j} \simeq \sum_{i=1}^N [\mathbfit{x}]_i \mathbfit{f}_i$) and testing the residual equation by appropriate test basis functions. This procedure leads to the definition of the linear system
\begin{equation}
    \eta \mat{T} \mathbfit{x} = \mat{e}^i
\end{equation}
where $\mat{T} = -\mathrm{j}k \, \mat{T}_s - (-\mathrm{j}k)^{-1} \, \mat{T}_h$, and the matrix entries are defined as $[\mat{T}_s]_{ij} = \langle \hat{\mathbf{n}} \times \mathbfit{f}_i, \mathcal{T}_{s,k} \mathbfit{f}_j \rangle$, $[\mat{T}_h]_{ij} = \langle \hat{\mathbf{n}} \times \mathbfit{f}_i, \mathcal{T}_{h,k} \mathbfit{f}_j \rangle$ and $[\mat{e}^i]_i = \langle \hat{\mathbf{n}} \times \mathbfit{f}_i, -\hat{\mathbf{n}} \times \mathbf{E}^i \rangle$, where $\langle \mathbfit{a}, \mathbfit{b} \rangle \coloneq \int_\Gamma \mathbfit{a}\cdot \mathbfit{b} \, \mathrm{d}S$. 
A typical choice for the basis functions $\mathbfit{f}_i$ is the set of Rao-Wilton-Glisson functions \cite{rao1982electromagnetic}.

This work builds upon the preconditioning strategies introduced in \cite{dely2019preconditioning} and focuses on the pseudo-inversion of the matrix obtained from the discretization of a shifted version of the Helmholtz operator, denoted as $\mathcal{H}$ and defined as
\begin{equation}
\label{eq:helmholtz_op}
\mathcal{H} = \Delta_\Gamma + k_m^2 \mathcal{I}\,.
\end{equation}
where $\Delta_\Gamma$ is the Laplace-Beltrami operator defined on $\Gamma$, $\mathcal{I}$ represents the identity operator, and $k_m = k - 0.4\mathrm{j}k^{1/3}d^{-2/3}$, with $d$ a characteristic dimension of the scatterer.
Its corresponding discretization $\mat{H_S}$ can be expressed as
\begin{equation}
    \mat{H_S} = -\mat{\Sigma}^T \mat{G}_{\mathbfit{\tilde{f}} \mathbfit{\tilde{f}}} \mat{\Sigma} + k_m^2 \mat{G}_{\tilde{\lambda} \tilde{\lambda}}
\end{equation}
where $\mat{\Sigma}$ is the transformation matrix from star to RWG subspaces, $\mat{G}_{\mathbfit{\tilde{f}} \mathbfit{\tilde{f}}}$ and $\mat{G}_{\tilde{\lambda} \tilde{\lambda}}$ are the Gram matrices ($[\mat{G}_{ab}]_{ij} = \int_\Gamma a_i \cdot b_j \mathrm{d} S$) associated with the Buffa-Christiansen and dual pyramid basis functions, denoted respectively as $\mathbfit{\tilde{f}}$ and $\mathbfit{\tilde{\lambda}}$. For further details on the background and notations, we refer the reader to \cite{dely2019preconditioning,adrian2021electromagnetic}.

In the proposed preconditioning strategy, we employ the single layer operator associated with the Green's kernel $G_{k_m}$,
\begin{equation}
(\mathcal{S}_{k_m}\psi)(\mathbfit{r}) = \int_\Gamma G_{k_m}(\mathbfit{r}, \mathbfit{r'}) \psi(\mathbfit{r'}) \, \mathrm{d}S'\,,
\end{equation}
discretized with patch basis functions $\{p_i\}_{i=1}^{N_c}(\mathbfit{r})$, where $N_c$ represents the number of cells of the mesh. This kind of discretization leads to the definition of matrix $\mat{S}$ with entry $[\mat{S}]_{ij} = \langle p_i, \mathcal{S}_{k_m} p_j \rangle$.
Moreover, the mixed Gram matrix $\mat{G}_{\tilde{\lambda}p}$ is employed to map the space of dual pyramid basis functions into the space of the patch ones.

\section{The Proposed Preconditioning}
\label{sec:new}
The main purpose of this work is to increase the computational efficiency of the $\mat{H_S}$ pseudo-inversion by applying a preconditioning strategy that regularizes the operator in all three aforementioned challenging regimes for certain classes of scatterers. 
The ill-conditioned nature of matrix $\mat{H_S}$ follows from the characteristics of the operator $\mathcal{H}$. In particular, the dense-discretization breakdown follows from the pseudo-differential order $+2$ of the Laplace operator, which leads to a condition number increase as $h^{-2}$ in regime (ii), where $h$ is the average mesh size.
On the other hand, the high-frequency breakdown is due to the spectral behavior of $\mathcal{H}$ at spectral indices $l \simeq (ka)$, where its eigenvalues, in magnitude, typically decrease with respect to the rest of the spectrum when increasing the frequency for a class of scatterers, causing the growth of the condition number of the matrix in regime (iii). The following analysis on the sphere confirms this statement (see \cref{eqn:Htrans}).

We propose to regularize matrix $\mat{H_S}$ by applying left- and right-multiplicative preconditioning matrices in the form
\begin{equation}
    \mat{H_S}_{\text{prec}} = \mat{S} \mat{G}_{\tilde{\lambda}p}^{-1}\mat{H_S}\mat{G}_{\tilde{\lambda}p}^{-\text{T}}\mat{S}\,.
\end{equation}
Indeed, to address the dense-discretization breakdown (i.e., regime ii), we exploit the fact that the single layer operator is of pseudo-differential order $-1$. This suggests that the preconditioned system $\mat{H_S}_{\text{prec}}$ should exhibit a mesh-independent conditioning. In fact, the global effect of the Helmholtz operator matrix and the single layer operator matrix on $\mat{H_S}$ is expected to be a clustering of the eigenvalues of the compound system. This occurs as a result of the pseudo-differential orders of the operators of the composition $\mathcal{S}_{k_m}\mathcal{H}\mathcal{S}_{k_m}$, respectively $-1$, $+2$ and $-1$, which sum up to $0$. 
Lastly, we notice that the application of the preconditioner $\mathcal{S}_{k_m}$ does not introduce spurious resonances into the system, thanks to the imaginary part of the complex wavenumber $k_m$.

To validate the effectiveness of the proposed formulation, a spherical harmonics analysis of the continuous operators defined over a sphere of radius $a$ is then performed. After denoting by $Y_l^m$ the spherical harmonic of order $l$ \cite{nationalinstituteofstandardsandtechnology2010nist}, it can be shown that \cite{nationalinstituteofstandardsandtechnology2010nist}
\begin{equation}
    \mathcal{H}^{k_m} Y_l^m = \left( -\frac{l(l+1)}{a^2} 
+ k_m^2 \right) Y_l^m
\end{equation}
and \cite{hsiao1994error}
\begin{equation}
    \mathcal{S}_{k_m} Y_l^m = -\mathrm{j} k_{m} a^2 j_l(k_{m}a) h_l^{(2)}(k_{m}a)\,,
\end{equation}
where $j_l$ and $h_l^{(2)}$ denote the spherical Bessel and Hankel functions \cite{nationalinstituteofstandardsandtechnology2010nist}.
Hence, by virtue of the commutability of the operators over the sphere, the eigenvalues of the composed operator $\mathcal{S}_{k_m}\mathcal{H}\mathcal{S}_{k_m}$ are
\begin{equation}
    \sigma(l,k) = k_{m}^2 a^4 \left[j_l(k_m a) h_l^{(2)}(k_{m}a)\right]^2\left[\frac{l(l+1)}{a^2}-k_m^2\right]\,.
\end{equation}

By substituting the asymptotic expansions for large order of the spherical Bessel and Hankel functions \cite[Eqn. 10.19(i)]{nationalinstituteofstandardsandtechnology2010nist}, we obtain that, for a fixed frequency,
\begin{equation}
    \lim_{l\to\infty}\sigma(l,k) = \lim_{l\to\infty} \frac{a^2 k_m^2-l(l+1)}{(2l+1)^2}
    =-\frac{1}{4}
\end{equation}
from which we infer that the spectrum of the preconditioned operator is bounded in the elliptic region of evanescent modes, suggesting the well-conditioning of the formulation with respect to the dense discretization parameter. 

In the hyperbolic region of propagative modes (for $l < ka$), instead, the eigenvalues of $\mathcal{S}_{k_m}$ decay in magnitude as $k^{-1}$ when the frequency increases \cite{buffa2006acoustic}, while it can be seen that the eigenvalues of $\mathcal{H}^{k_m}$ increase as $k^2$, from which we infer the boundedness of the eigenvalues of the preconditioned operator in this region. Moreover, the complex wavenumber employed leads to the absence of spurious resonances in the spectrum of $\mathcal{S}^{k_m}$, preventing the introduction of quasi-null eigenvalues in the spectrum of the preconditioned operator. 

\begin{figure}
    \centering
    \includegraphics[width=0.9\columnwidth]{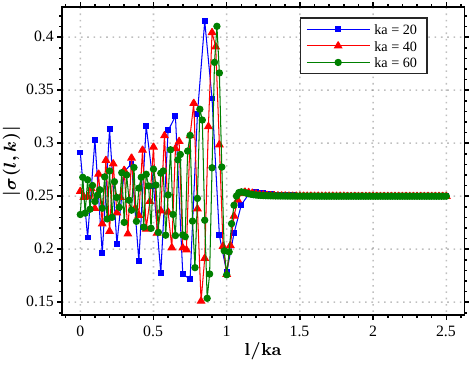}
    \caption{Comparison of the spectra of the $\mathcal{S}_{k_m} \mathcal{H} \mathcal{S}_{k_m}$ operator evaluated for different $ka$ values via Spherical Harmonics.}
    \label{fig:SHS_SH_spectra.pdf}
\end{figure}

Finally, in the transition region, characterized by spectral indices $l\simeq (ka)$, the eigenvalues of $\mathcal{S}^{k_m}$ decay in frequency as $k^{-2/3}$ \cite{buffa2006acoustic}, while the eigenvalues of $\mathcal{H}^{k_m}$ increase in magnitude as $k^{4/3}$. Indeed, for $k\to \infty$,
\begin{align}
    \frac{-ka(ka+1)}{a^2}&+(k-\mathrm{j}0.4k^{1/3}a^{-2/3})^2 = \nonumber\\&\quad\quad\quad\quad -2\mathrm{j}0.4k^{4/3}a^{-2/3}+\mathcal{O}(k)\,,
    \label{eqn:Htrans}
\end{align}
so that the spectral behavior of the preconditioned operator is constant in frequency.
These considerations are corroborated by results in \autoref{fig:SHS_SH_spectra.pdf}, which shows the magnitude of the eigenvalues of the preconditioned operator for diverse frequencies.
Similar spectral trends are expected for smooth convex geometries other than the sphere.

So, when applied to these classes of scatterers, the proposed approach stabilizes the condition number of the system in the dense-discretization (ii) and high-frequency (iii) regimes; for $k \to 0$ it is worth noting that $\mathcal{S}_{k_m}\mathcal{H}\mathcal{S}_{k_m}$ converges to the well-conditioned form $\mathcal{S}_0 \Delta_\Gamma \mathcal{S}_0$ \cite{oneil2018secondkind}, where $\mathcal{S}_0$ represents the single layer operator associated with the static Green's function kernel.

\section{Numerical Results}
\label{sec:results}
To corroborate the theoretical results, numerical tests have been performed on a sphere of radius one ($a=\SI{1}{\meter}$) discretized with triangular patches obtained as the result of a recursive subdivision of an icosahedral mesh. 
For simulation purposes, a randomly generated right hand side (RHS) vector was built and then, to be consistent with the physics of the formulation, orthogonalized against the nullspace of the Laplacian operator (i.e., the constant vector).
The systems were solved using the built-in GMRES algorithm in MATLAB with a restart parameter of 100 and a relative residual tolerance of $10^{-6}$.
The behavior of the standard shifted Helmholtz operator $\mat{H_S}$ was compared against that of the preconditioned system $\mat{H_S}_{\text{prec}}$ in the two challenging regimes analyzed in this work.

\subsection{Dense-discretization behavior}

\begin{figure}
    \centering
    \includegraphics[width=0.9\columnwidth]{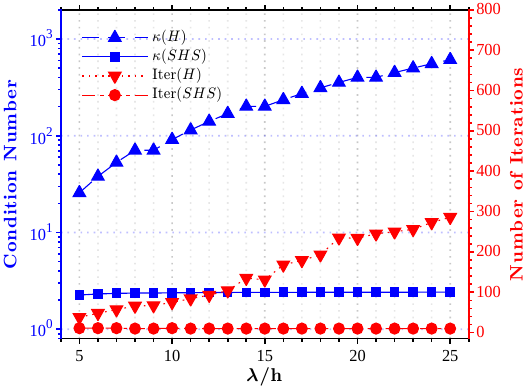}
    \caption{Comparison between $\mat{H_S}$ and $\mat{H_S}_{\text{prec}}$ in terms of condition number and GMRES iterations when varying the mesh density $\lambda/h$. The preconditioned system exhibits stability against mesh refinement.}
    \label{fig:increase_loh.pdf}
\end{figure}

In the first experiment, focus was placed on how the condition number ($\kappa(M)$) of the system and the number of iterations ($Iter(M)$) required by GMRES to reach convergence change while the discretization is refined maintaining the frequency constant.
Hence, we fixed the wavenumber $k=\SI{4}{\per\meter}$, corresponding to a frequency of $\SI{191}{\mega\hertz}$, and varied the $\lambda / h$ ratio from $5$ to $25$ in increments of $0.5$. 
According to the theoretical analysis, \autoref{fig:increase_loh.pdf} exhibits a stable trend of the condition number of the preconditioned system $\mat{H_S}_{\text{prec}}$ while decreasing the mesh size. This correlates with a quasi-constant number of iterations required by the iterative solver. On the other hand, the condition number and the number of iterations for the original shifted Helmholtz operator increase sharply as the $\lambda / h$ ratio rises, revealing the known ill-conditioning problem of the original system.

\subsection{High-frequency behavior}
\begin{figure}
    \centering
    \includegraphics[width=0.9\columnwidth]{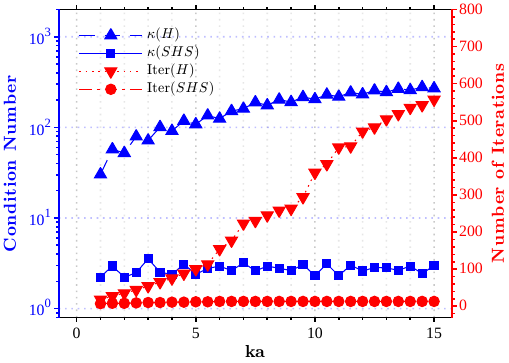}
    \caption{Comparison between $\mat{H_S}$ and $\mat{H_S}_{\text{prec}}$ in terms of condition number and GMRES iterations when varying the electrical size $ka$. The proposed preconditioner effectively suppresses the high-frequency breakdown and does not introduce spurious resonances.}
    \label{fig:increase_k.pdf}
\end{figure}

The second set of experiments focuses on the high-frequency regime. The $\lambda / h$ ratio was fixed at $10$ while $k$ varied from $\SI{1}{\per\meter}$ to $\SI{15}{\per\meter}$ in increments of $\SI{1}{\per\meter}$. The results illustrated in \autoref{fig:increase_k.pdf} show that the number of iterations and the condition number of  the preconditioned system are stable, while the corresponding quantities for $\mat{H_S}$ keep increasing as $ka$ rises.

\section{Conclusion}
A preconditioning strategy to improve the Helmholtz operator pseudo-inversion and, consequently, to accelerate the solution of the preconditioned EFIE has been explored. As corroborated by both theoretical and numerical results, the single layer operator, equipped with a complex wavenumber kernel, successfully stabilizes the resulting system in the dense-discretization and high-frequency regimes.

While this work focuses on the theoretical validation on canonical geometries, future works will assess the computational resource requirements and the performance of the approach for a large class of practically relevant scatterers, as well as leverage fast algorithms such as the FMM to accelerate the pseudo-inversion of the preconditioned matrix.

\section*{ACKNOWLEDGEMENT}
\ifshowinfo
This work has received funding from the European Innovation Council (EIC) through the European Union’s Horizon Europe research Programme under Grant 101046748 (Project CEREBRO).
\fi

{ \printbibliography }

\end{document}